%% file: main.tex
\documentclass[conference]{IEEEtran}
\IEEEoverridecommandlockouts
\usepackage{cite}
\usepackage{amsmath,amssymb,amsfonts}
\usepackage{algorithmic}
\usepackage{graphicx}
\usepackage{textcomp}
\usepackage{xcolor}
\usepackage{url}
\usepackage{multirow} 
\input{acronyms}

\def\BibTeX{{\rm B\kern-.05em{\sc i\kern-.025em b}\kern-.08em
    T\kern-.1667em\lower.7ex\hbox{E}\kern-.125emX}}
\begin{document}

\title{TARNet: A Temporal-Aware Multi-Scale Architecture for Closed-Set Speaker Identification\thanks{Accepted for publication at the IEEE International Conference on Multimedia and Expo (ICME) 2026. This is the authors' accepted manuscript version prepared for arXiv. The final published version, when available, will appear in IEEE Xplore.}}

\author{
\IEEEauthorblockN{Yassin TERRAF\IEEEauthorrefmark{1}\IEEEauthorrefmark{2}, Youssef IRAQI\IEEEauthorrefmark{1}}
\IEEEauthorblockA{\IEEEauthorrefmark{1}College of Computing, University Mohammed VI Polytechnic, Benguerir, Morocco}
\IEEEauthorblockA{\IEEEauthorrefmark{2}CID Development, Rabat, Morocco}
\IEEEauthorblockA{Email: yassin.terraf@um6p.ma, youssef.iraqi@um6p.ma}
}
\maketitle

\begin{abstract}

Closed-set speaker identification aims to assign a speech utterance to one of a predefined set of enrolled speakers and requires robust modeling of speaker-specific characteristics across multiple temporal scales. While recent deep learning approaches have achieved strong performance, many existing architectures provide limited mechanisms for modeling temporal dependencies across different time scales, which can restrict the effective use of complementary short-, mid-, and long-term speaker characteristics.
In this paper, we propose TARNet, a lightweight Temporal-Aware Representation Network for closed-set speaker identification. TARNet explicitly models temporal information at multiple time scales using a multi-stage temporal encoder with stage-specific dilation configurations. The resulting multi-scale representations are fused and aggregated via an Attentive Statistics Pooling (ASP) module to produce a discriminative utterance-level speaker embedding. Experiments on the VoxCeleb1 and LibriSpeech datasets show that TARNet outperforms state-of-the-art methods while maintaining competitive computational complexity, making it suitable for practical speaker identification systems. The code is publicly available at \url{https://github.com/YassinTERRAF/TARNet}.

\end{abstract}


\begin{center}
\footnotesize\textit{Accepted for publication at IEEE ICME 2026.}
\end{center}

\glsresetall

\section{Introduction}
\label{sec:introduction}

Closed-set speaker identification aims to determine the identity of a speaker from a given speech utterance by assigning it to one of a predefined set of enrolled speakers~\cite{10410836}. This task plays an important role in applications such as biometric authentication~\cite{singh2018voice} and forensic analysis~\cite{GOURI2024110074}.
Early speaker identification systems relied on handcrafted acoustic features combined with \gls{dnn}. Jahangir et al.~\cite{8995509} employed \gls{mfcc}, while Rahman et al.~\cite{9631309} explored prosodic features for DNN-based speaker classification. Salvati et al.~\cite{SALVATI2023119750} later proposed a hybrid representation combining raw waveform and gammatone cepstral features processed by parallel DNN branches. Although effective, these DNN-based approaches apply fully connected networks to individual speech frames and therefore do not model speaker-related temporal dependencies across an utterance.

To address the limitations of \gls{dnn}s in modeling speech representations, subsequent work explored convolutional neural network \gls{cnn}-based architectures to learn more robust spectral features. Among \gls{cnn}-based approaches, VGG-style architectures, adapted from the VGG-M network originally developed for image recognition, have been widely adopted for speaker recognition~\cite{NAGRANI2020101027}. Nagrani et al.~\cite{NAGRANI2020101027} employed VGG-M with spectrogram features, while Chung et al.~\cite{chung20_odyssey} adapted the architecture to log-Mel filterbank representations. Extensions such as VGG-\gls{cnn}~\cite{HAMSA2023119871}, which integrate a masking-based front-end for noise-robust feature learning with a SpeechVGG network, further improve speaker identification performance under diverse recording conditions. Anidjar et al.~\cite{ANIDJAR2024124671} proposed DLSI-SM-VGG-M, which builds upon the VGG-M architecture while reducing architectural complexity through the use of smaller convolutional kernels and a double log-softmax loss function, and demonstrates strong performance for speaker identification. Despite the strong performance of \gls{cnn}-based architectures for speaker identification, temporal modeling is typically achieved implicitly through network depth and stacked convolutions. Consequently, speaker-related information from different temporal scales is often merged into a single frame-level representation, which limits the model’s ability to explicitly capture and leverage short-, mid-, and long-term speaker characteristics. In addition, the effective temporal receptive field of \gls{cnn}-based models is often limited, which can restrict the capture of long-term speaker characteristics. Moreover, temporal aggregation of frame-level features commonly relies on simple pooling operations, which assign equal importance to all frames.

To address these challenges, we propose TARNet, a temporal-aware representation network for speaker identification. TARNet consists of a multi-scale temporal encoder built from \gls{tcn} blocks, a multi-scale feature fusion module, an \gls{asp} layer, and a classification head. The temporal encoder is organized into multiple stages with scale-specific dilation patterns and repeated \gls{tcn} blocks, enabling explicit modeling of short-, mid-, and long-term temporal dependencies while preserving temporal resolution. The resulting multi-scale representations are fused and aggregated using \gls{asp}, which assigns different importance to frames to produce a compact utterance-level speaker representation.

\begin{figure*}[!ht]
\centering
\includegraphics[width=\textwidth]{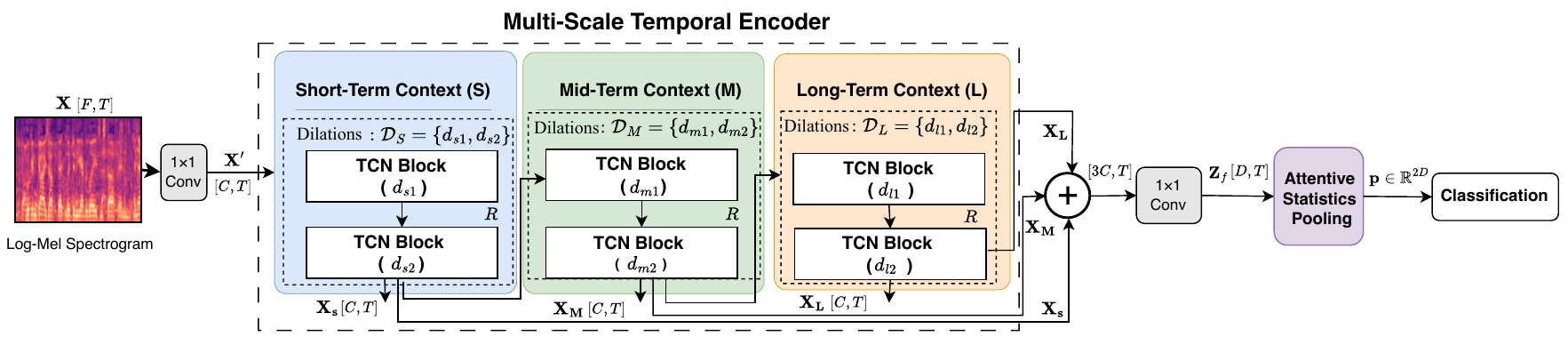}
\caption{The proposed TARNet architecture for speaker identification. The network consists of an acoustic front-end with bottleneck projection, a multi-scale temporal encoder, an \gls{asp} layer, and a final classification module. The $\oplus$ denotes channel-wise concatenation of $\mathbf{X}_S$, $\mathbf{X}_M$, and $\mathbf{X}_L$.}

\label{fig:Tarnet_architecture}
\end{figure*}

The main contributions of this work are as follows:
\begin{itemize}
    \item We propose TARNet, a lightweight architecture for closed-set speaker identification that explicitly models speaker information at multiple temporal scales.
    
    \item We investigate the impact of handcrafted and self-supervised speech representations within the proposed approach for speaker identification.
    
    \item We demonstrate through extensive experiments on the VoxCeleb1 and LibriSpeech datasets that TARNet outperforms state-of-the-art speaker identification baselines.
    
    \item We evaluate the computational efficiency of TARNet and baselines in terms of model size and inference time.
    
\end{itemize}

The remainder of this paper is organized as follows. Section~\ref{sec:tarnet_architecture} describes the proposed TARNet architecture. Section~\ref{sec:experiments} presents the experimental setup and evaluation protocol. The experimental results and ablation studies are reported in Section~\ref{sec:experiments_results}. Finally, Section~\ref{sec:conclusion} concludes the paper.

\section{TARNet Network Architecture}
\label{sec:tarnet_architecture}

This section presents the TARNet architecture. As shown in Fig.~\ref{fig:Tarnet_architecture}, TARNet consists of an acoustic front-end with bottleneck projection, a multi-scale temporal encoder, an attentive aggregation stage, and a final classification module.

\subsection{Feature Extraction and Bottleneck Projection}
\label{subsec:feature_extraction_bottleneck}

Effective acoustic feature extraction is essential for representing speaker-dependent characteristics in speech signals. In this work, log-Mel spectrogram features are extracted following standard time--frequency processing, including short-time Fourier transform, Mel filterbank projection, and logarithmic compression. The resulting representation is denoted by $\mathbf{X} \in \mathbb{R}^{F \times T}$, where $F$ is the number of Mel frequency bands and $T$ is the number of time frames. To facilitate subsequent temporal modeling, a bottleneck projection is applied using a $1\times1$ convolution, which linearly projects $\mathbf{X}$ into a compact channel representation $\mathbf{X}' \in \mathbb{R}^{C \times T}$, where $C$ denotes the number of output channels.

\subsection{Multi-Scale Temporal Encoder}
\label{subsec:multiscale_encoder}

\begin{figure}[!ht]
\centering
\includegraphics[width=\columnwidth]{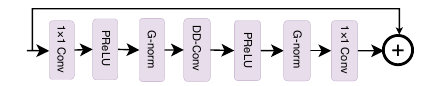}
\caption{The details of the TCN block. The ``DD-Conv'' indicates a dilated depthwise separable convolution. The ``G-norm'' refers to global layer normalization.}
\label{fig:tcn}
\end{figure}

In speaker identification, frame-level features must be integrated over time to form an utterance-level representation for classification. Temporal modeling is therefore required to capture speaker-related characteristics that manifest at multiple time scales, ranging from short-term acoustic patterns to longer-term speaking characteristics~\cite{10.1007/978-3-032-02548-7_7}. To address this, TARNet incorporates a multi-scale temporal encoder, as illustrated in Fig.~\ref{fig:Tarnet_architecture}. The encoder operates on the bottleneck features $\mathbf{X}' \in \mathbb{R}^{C \times T}$ and is organized into three cascaded temporal stages. Each stage captures speaker-related information over a specific temporal range while progressively expanding the temporal context and receptive field without reducing temporal resolution. Each stage consists of stacked \gls{tcn} blocks~\cite{lea2016temporal}, which model temporal dependencies using dilated one-dimensional convolutions along the time axis. \gls{tcn}s enable efficient modeling of multi-scale temporal context while preserving temporal resolution, making them well suited for capturing speaker-related temporal characteristics. As illustrated in Fig.~\ref{fig:tcn}, a \gls{tcn} block comprises a pointwise $1\times1$ convolution for channel mixing, followed by a depthwise dilated one-dimensional convolution along the time axis to model temporal dependencies at a scale determined by the dilation factor. Furthermore, PReLU nonlinear activation and normalization layers are applied between convolutional operations, and a residual connection links the block input and output to facilitate stable optimization.

Let $\mathbf{X}_{0}=\mathbf{X}' \in \mathbb{R}^{C \times T}$ denote the input to the multi-scale temporal encoder, where $C$ is the number of channels and $T$ is the number of time frames. Each temporal stage is composed of \gls{tcn} blocks, denoted by $\mathcal{B}_{d}(\cdot)$, where $d$ represents the dilation factor. Each TCN block preserves the temporal resolution and channel dimensionality, thereby enabling residual connections and consistent feature propagation across stages. The encoder employs three cascaded, stage-specific dilation sets, $\mathcal{D}_{S}$, $\mathcal{D}_{M}$, and $\mathcal{D}_{L}$, corresponding to short-, mid-, and long-term temporal contexts, respectively. Within each stage, the sequence of dilation factors is repeated $R$ times, where $R$ denotes the number of repetitions, which increases the effective receptive field and enables multiple nonlinear transformations at a fixed temporal scale, resulting in more refined temporal representations for the corresponding temporal range. 

The first stage focuses on short-term temporal context and consists of TCN blocks with dilation set $\mathcal{D}_{S}=\{d_{s,1}, d_{s,2}\}$, where $d_{s,1}$ and $d_{s,2}$ denote the dilation factors used in this stage. The stage output is computed by repeating the corresponding TCN block sequence $R$ times as follows:
\begin{equation}
\mathbf{X}_{S} = \left(\mathcal{B}_{d_{s,2}} \circ \mathcal{B}_{d_{s,1}}\right)^{R} (\mathbf{X}_{0}).
\end{equation}
Here, $\circ$ denotes function composition, and $(\cdot)^R$ indicates $R$ successive repetitions of the composed TCN blocks.

Building on $\mathbf{X}_{S}$, the second stage captures mid-term temporal context using a larger dilation set $\mathcal{D}_{M}=\{d_{m,1}, d_{m,2}\}$:
\begin{equation}
\mathbf{X}_{M} =
\left(\mathcal{B}_{d_{m,2}} \circ \mathcal{B}_{d_{m,1}}\right)^{R}
(\mathbf{X}_{S}).
\end{equation}

The third stage further extends the temporal context by operating on $\mathbf{X}_{M}$ with the largest dilation set $\mathcal{D}_{L}=\{d_{l,1}, d_{l,2}\}$, enabling the modeling of long-term temporal dependencies:
\begin{equation}
\mathbf{X}_{L} =
\left(\mathcal{B}_{d_{l,2}} \circ \mathcal{B}_{d_{l,1}}\right)^{R}
(\mathbf{X}_{M}).
\end{equation}

The resulting representations $\mathbf{X}_{S}$, $\mathbf{X}_{M}$, and $\mathbf{X}_{L}$ capture speaker-related features at short-, mid-, and long-term temporal scales, respectively. These representations are concatenated along the channel dimension and projected using a $1\times1$ convolution to form a unified multi-scale representation:
\begin{equation}
\mathbf{Z}_{f} =
\phi\!\left(\mathrm{Conv}_{1\times1}
\big([\mathbf{X}_{S} \, \| \, \mathbf{X}_{M} \, \| \, \mathbf{X}_{L}]\big)\right)
\in \mathbb{R}^{D \times T},
\end{equation}
where $[\cdot \| \cdot]$ denotes channel-wise concatenation, $D$ is the number of
output channels, and $\phi(\cdot)$ denotes the ReLU activation function.

\subsection{Attentive Statistics Pooling}
\label{subsec:asp}

In speaker identification, speaker-discriminative information is unevenly distributed over time, and different speech frames contribute unequally to speaker representation~\cite{10.1007/978-981-96-6594-5_6}. To address this, TARNet employs the standard \gls{asp} module proposed by Okabe et al.~\cite{Okabe_ASP} to aggregate the fused frame-level representation $\mathbf{Z}_{f} \in \mathbb{R}^{D \times T}$ into a fixed-dimensional utterance-level embedding. ASP assigns channel-dependent attention weights to frame-level features, enabling the model to emphasize frames that are more informative for speaker discrimination.

Let $\mathbf{z}(t) \in \mathbb{R}^{D}$ denote the fused feature vector at time index $t$. An attention context vector $\mathbf{c}(t)$ is constructed by concatenating the frame-level feature with the global mean and standard deviation computed over the temporal dimension:
\begin{equation}
\mathbf{c}(t) =
\left[\mathbf{z}(t)\; \| \; \boldsymbol{\mu} \; \| \; \boldsymbol{\sigma}\right]
\in \mathbb{R}^{3D},
\end{equation}
where $\boldsymbol{\mu} \in \mathbb{R}^{D}$ and $\boldsymbol{\sigma} \in \mathbb{R}^{D}$ denote the global mean and standard deviation of $\mathbf{z}(t)$ across all time frames, respectively.

The attention weights are predicted using a lightweight two-layer $1\times1$ convolutional network that takes $\mathbf{c}(t)$ as input. The network applies a nonlinear transformation followed by a softmax operation along the temporal dimension to produce normalized, channel-dependent attention weights $\boldsymbol{\alpha}(t) \in \mathbb{R}^{D}$, where each channel is assigned an independent attention weight at each time frame.

Using these attention weights, \gls{asp} computes weighted first- and second-order statistics of the frame-level features:
\begin{equation}
\boldsymbol{\mu}_{a} = \sum_{t=1}^{T}\boldsymbol{\alpha}(t)\odot\mathbf{z}(t),
\end{equation}
\begin{equation}
\boldsymbol{\sigma}_{a} =
\sqrt{\sum_{t=1}^{T}\boldsymbol{\alpha}(t)\odot\mathbf{z}(t)^2 -
\boldsymbol{\mu}_{a}^2},
\end{equation}
where $\odot$ denotes element-wise multiplication.

The final utterance-level representation is obtained by concatenating the weighted mean and standard deviation:
\begin{equation}
\mathbf{p} =
\left[\boldsymbol{\mu}_{a} \; \| \; \boldsymbol{\sigma}_{a}\right]
\in \mathbb{R}^{2D}.
\end{equation}

The utterance-level representation $\mathbf{p}$ is subsequently projected to a compact speaker embedding and passed to a linear classifier with softmax activation to produce speaker class probabilities.

\begin{table*}[!ht]
\centering
\caption{Experimental results on the VoxCeleb1 test set. Results are reported as
mean $\pm$ standard deviation (\%). Best results are shown in bold, and
second-best results are underlined.}
\label{table:clean_results_voxceleb}
\small
\begin{tabular}{lccccc}
\hline
\textbf{Model} & \textbf{Top-1 Acc.} & \textbf{Top-5 Acc.} & \textbf{Precision} & \textbf{Recall} & \textbf{F1-score} \\
\hline
\gls{cnn}-no-norm        & $67.59 \pm 0.10$ & $85.30 \pm 0.06$ & $67.81 \pm 0.08$ & $67.59 \pm 0.07$ & $67.70 \pm 0.03$ \\
VGG-M                    & $80.81 \pm 0.06$ & $92.93 \pm 0.03$ & $80.63 \pm 0.03$ & $80.81 \pm 0.01$ & $80.72 \pm 0.07$ \\
VGG-M-40                 & $73.11 \pm 0.02$ & $88.96 \pm 0.07$ & $73.26 \pm 0.03$ & $73.11 \pm 0.08$ & $73.18 \pm 0.03$ \\
ResNetSE-34L             & $77.45 \pm 0.03$ & $90.17 \pm 0.07$ & $77.18 \pm 0.09$ & $77.45 \pm 0.09$ & $77.31 \pm 0.03$ \\
Thin ResNet-34           & $86.85 \pm 0.06$ & $96.22 \pm 0.04$ & $86.97 \pm 0.03$ & $86.85 \pm 0.07$ & $86.91 \pm 0.07$ \\
DCSA-ResNet18            & $81.88 \pm 0.06$ & $93.54 \pm 0.02$ & $81.68 \pm 0.06$ & $81.88 \pm 0.07$ & $81.78 \pm 0.04$ \\
VGG-\gls{cnn}            & $86.14 \pm 0.08$ & $95.76 \pm 0.04$ & $86.40 \pm 0.03$ & $86.14 \pm 0.07$ & $86.27 \pm 0.02$ \\
ResNeXt                  & $88.83 \pm 0.02$ & $96.81 \pm 0.03$ & $88.69 \pm 0.02$ & $88.83 \pm 0.06$ & $88.76 \pm 0.04$ \\
LF-DNN-GCC               & $83.34 \pm 0.05$ & $94.12 \pm 0.04$ & $83.53 \pm 0.03$ & $83.34 \pm 0.06$ & $83.43 \pm 0.07$ \\
DLSI-SM-VGG-M            & $90.04 \pm 0.05$ & $97.20 \pm 0.06$ & $89.79 \pm 0.04$ & $90.04 \pm 0.05$ & $89.91 \pm 0.07$ \\
\hline
x-vector                 & $91.89 \pm 0.07$ & $97.67 \pm 0.01$ & $92.06 \pm 0.07$ & $91.89 \pm 0.07$ & $91.97 \pm 0.02$ \\
ECAPA-TDNN               & \underline{$94.50 \pm 0.07$} & \underline{$98.32 \pm 0.04$} & \underline{$94.29 \pm 0.03$} & \underline{$94.50 \pm 0.04$} & \underline{$94.39 \pm 0.04$} \\
TARNet & 
$\mathbf{96.25} \pm \mathbf{0.04}$ &
$\mathbf{98.91} \pm \mathbf{0.02}$ &
$\mathbf{96.49} \pm \mathbf{0.07}$ &
$\mathbf{96.25} \pm \mathbf{0.10}$ &
$\mathbf{95.78} \pm \mathbf{0.02}$ \\
\hline
\end{tabular}
\end{table*}

\section{Experimental Protocol}
\label{sec:experiments}

\subsection{Datasets}
\label{sec:datasets}

\subsubsection{VoxCeleb Dataset}
\label{sec:voxceleb}

VoxCeleb~\cite{NAGRANI2020101027} is a large-scale audio-visual dataset collected from YouTube interviews, widely used for speaker identification and verification. It contains short speech segments from diverse speakers under unconstrained conditions with variability in noise, channels, and environments. The dataset includes 1,251 speakers (690 male and 561 female) and 153,516 utterances.

\subsubsection{LibriSpeech Dataset}
\label{sec:librispeech}

LibriSpeech~\cite{7178964} is a public speech corpus derived from LibriVox audiobooks, widely used in speech and speaker recognition. It contains about 1,000 hours of 16 kHz read English speech. We use the \textit{train-clean-100} subset, which includes 251 speakers (126 male and 125 female) and 28,539 utterances.

\subsection{Experimental Settings}
\label{sec:experimental_settings}

For TARNet, 80-dimensional log-Mel spectrogram features are used as acoustic input. These features are compared with self-supervised representations in the ablation study (Section~\ref{sec:ablation}). For the VoxCeleb dataset, we follow the official closed-set speaker identification protocol defined in~\cite{NAGRANI2020101027}. During training, speech segments are randomly cropped to 3-second segments, while full-length utterances are used during inference. For the LibriSpeech dataset, utterances are randomly split into 70\% training, 10\% validation, and 20\% testing, and further segmented into fixed-length 2-second speech segments, from which log-Mel features are extracted using the same procedure. For fair comparison, all baseline models are trained under the same training protocol as TARNet, while architecture-specific hyperparameters are independently tuned based on validation performance. In the multi-scale temporal encoder, dilation factors are set to $\{1,2\}$, $\{4,8\}$, and $\{16,32\}$ for the short-, mid-, and long-term stages, respectively, with each stage repeated $R=3$ times. These hyperparameters are selected to balance classification performance and computational complexity. All models are trained using stochastic gradient descent with an initial learning rate of 0.001 and a weight decay of $5\times10^{-4}$. Training is performed for 300 epochs with a batch size of 100. Experiments are conducted on an Intel Xeon CPU and an NVIDIA A100 GPU with 80\,GB of memory.

\subsection{Evaluation Metrics}
\label{sec:evaluation_metrics}

The proposed TARNet architecture is evaluated using standard speaker identification metrics, including Top-1 and Top-5 accuracy, weighted precision, weighted recall, and weighted F1-score. Top-1 accuracy measures whether the correct speaker is the highest-ranked prediction, while Top-5 accuracy evaluates whether the correct speaker appears among the five highest-ranked predictions. To assess the statistical significance of performance differences between models, we employ the \gls{ar} test~\cite{noreen1989computer}, a non-parametric permutation-based significance test.

\subsection{Baselines}
\label{sec:baselines}
To evaluate the effectiveness of the proposed TARNet, we compare it with representative and state-of-the-art speaker identification models. These include \gls{cnn}- and ResNet-based architectures such as \gls{cnn}-no-norm~\cite{NAGRANI2020101027}, VGG-M~\cite{NAGRANI2020101027,1467314} and its modified variant~\cite{chung20_odyssey}, Thin ResNet-34~\cite{chung20_odyssey}, ResNetSE-34L~\cite{interspechFastResnet34SE}, DCSA-ResNet18~\cite{an2019deep}, VGG-\gls{cnn}~\cite{HAMSA2023119871}, ResNeXt~\cite{9383531}, LF-DNN-GCC~\cite{SALVATI2023119750}, and DLSI-SM-VGG-M~\cite{ANIDJAR2024124671}.
We also include TDNN-based models that explicitly capture temporal dependencies across speech frames, namely x-vector~\cite{8461375} and ECAPA-TDNN~\cite{desplanques2020ecapa}. These models are adapted for closed-set speaker identification by replacing the verification backend with a classification head. The fixed-dimensional embeddings produced by x-vector and ECAPA-TDNN are passed through a ReLU activation followed by a linear classification layer to predict speaker classes.

\section{Experimental Results}
\label{sec:experiments_results}

This section presents the experimental evaluation of TARNet, including comparisons with baseline methods on VoxCeleb1 and LibriSpeech, ablation studies, and an analysis of computational complexity.

\begin{table*}[!ht]
\centering
\caption{Experimental results on the LibriSpeech test set. Results are
reported as mean $\pm$ standard deviation (\%). Best results are shown in bold,
and second-best results are underlined.}
\label{table:librispeech_clean_avs}
\small
\begin{tabular}{lccccc}
\hline
\textbf{Model} & \textbf{Top-1 Acc.} & \textbf{Top-5 Acc.} & \textbf{Precision} & \textbf{Recall} & \textbf{F1-score} \\
\hline
\gls{cnn}-no-norm   & $90.33 \pm 0.77$ & $98.03 \pm 0.45$ & $90.54 \pm 0.75$ & $90.32 \pm 0.75$ & $90.43 \pm 0.76$ \\
VGG-M               & $92.99 \pm 0.62$ & $98.47 \pm 0.02$ & $92.80 \pm 0.62$ & $92.98 \pm 0.62$ & $92.89 \pm 0.64$ \\
VGG-M-40            & $93.38 \pm 0.51$ & $98.38 \pm 0.35$ & $93.51 \pm 0.52$ & $93.36 \pm 0.52$ & $93.43 \pm 0.52$ \\
ResNetSE-34L        & $91.56 \pm 0.45$ & $98.40 \pm 0.12$ & $91.28 \pm 0.45$ & $91.55 \pm 0.45$ & $91.41 \pm 0.41$ \\
Thin ResNet-34      & $97.36 \pm 0.11$ & $99.58 \pm 0.05$ & $97.48 \pm 0.10$ & $97.36 \pm 0.10$ & $97.42 \pm 0.09$ \\
DCSA-ResNet18       & $96.40 \pm 0.08$ & $99.65 \pm 0.05$ & $96.19 \pm 0.07$ & $96.39 \pm 0.07$ & $96.29 \pm 0.08$ \\
VGG-\gls{cnn}       & $97.10 \pm 0.02$ & $99.51 \pm 0.10$ & $97.36 \pm 0.02$ & $97.10 \pm 0.02$ & $97.23 \pm 0.04$ \\
ResNeXt             & $97.24 \pm 0.17$ & $99.45 \pm 0.05$ & $97.10 \pm 0.17$ & $97.24 \pm 0.17$ & $97.17 \pm 0.19$ \\
LF-DNN-GCC          & $93.84 \pm 0.36$ & $98.49 \pm 0.09$ & $94.02 \pm 0.37$ & $93.83 \pm 0.37$ & $93.92 \pm 0.35$ \\
DLSI-SM-VGG-M       & $97.52 \pm 0.15$ & $99.62 \pm 0.06$ & $97.27 \pm 0.15$ & $97.52 \pm 0.15$ & $97.39 \pm 0.16$ \\
\hline
x-vector            & $93.23 \pm 0.41$ & $98.42 \pm 0.00$ & $93.39 \pm 0.42$ & $93.22 \pm 0.42$ & $93.30 \pm 0.41$ \\
ECAPA-TDNN          & \underline{$97.80 \pm 0.19$} & \underline{$99.68 \pm 0.08$} & \underline{$97.59 \pm 0.17$} & \underline{$97.80 \pm 0.17$} & \underline{$97.69 \pm 0.18$} \\
TARNet &
$\boldsymbol{99.25 \pm 0.07}$ &
$\boldsymbol{99.74 \pm 0.02}$ &
$\boldsymbol{99.48 \pm 0.07}$ &
$\boldsymbol{99.24 \pm 0.07}$ &
$\boldsymbol{99.36 \pm 0.08}$ \\
\hline
\end{tabular}
\end{table*}

\subsection{Comparative Experiments on the VoxCeleb Dataset}

Table~\ref{table:clean_results_voxceleb} reports the closed-set speaker identification results on the VoxCeleb1 test set. TARNet achieves the best performance across all reported metrics, reaching a Top-1 accuracy of $96.25\%$ and a Top-5 accuracy of $98.91\%$. Compared with the strongest baseline, ECAPA-TDNN, TARNet improves Top-1 accuracy by 1.75 percentage points, with the difference being statistically significant according to the \gls{ar} test. Relative to the x-vector baseline, TARNet achieves a larger and statistically significant Top-1 accuracy gain of 4.36 points, indicating that the proposed architecture remains effective even when compared with strong embedding extractors adapted to the identification setting.

When compared with \gls{cnn}-based speaker identification baselines, TARNet significantly outperforms the strongest competing method, DLSI-SM-VGG-M ($90.04\%$), by 6.21 percentage points in Top-1 accuracy.

\subsection{Comparative Experiments on the LibriSpeech Dataset}

Table~\ref{table:librispeech_clean_avs} reports closed-set speaker identification results on the LibriSpeech test set. Due to the controlled recording conditions, most baseline methods already achieve strong performance, making further improvements challenging. Despite this, TARNet consistently achieves the best results across all evaluation metrics, with a Top-1 accuracy of $99.25\%$ and a Top-5 accuracy of $99.74\%$. Compared with the strongest baseline, ECAPA-TDNN, TARNet yields an absolute improvement of $1.45$ points in Top-1 accuracy. Among \gls{cnn}-based speaker identification models, Thin ResNet-34 and DLSI-SM-VGG-M also demonstrate strong performance, achieving over $97\%$ Top-1 accuracy.

These improvements are attributed to the explicit modeling of speaker-related temporal features at multiple time scales. The multi-scale temporal encoder captures both short- and long-term speaker characteristics while preserving intermediate representations prior to fusion, thereby reducing the loss of discriminative temporal information. In addition, \gls{asp} emphasizes speaker-discriminative frames and suppresses less informative segments. As a result, TARNet achieves consistent performance gains on both VoxCeleb1, which features unconstrained and realistic recording conditions, and LibriSpeech, which represents a clean and controlled evaluation setting.

\subsection{Ablation Study}
\label{sec:ablation}

\subsubsection{Feature Extraction}
\label{sec:ablation_features}

This ablation study examines the impact of the input feature representation on TARNet performance. We compare handcrafted log-Mel spectrogram features with widely used self-supervised learning (SSL) speech representations, including WavLM, wav2vec~2.0, and HuBERT, using publicly available Base pretrained models\footnote{\url{https://huggingface.co/microsoft/wavlm-base}}
\footnote{\url{https://huggingface.co/facebook/wav2vec2-base}}
\footnote{\url{https://huggingface.co/facebook/hubert-base-ls960}}.
Two SSL configurations are evaluated. In the first setting, all SSL model parameters are frozen and frame-level embeddings are extracted from the final transformer encoder layer. In the second setting, only the final transformer encoder layer is fine-tuned, while all preceding layers remain frozen.

\begin{table}[!ht]
\centering
\caption{Ablation study on input feature representations evaluated on the
VoxCeleb1 test set.}
\label{table:ablation_features}
\small
\begin{tabular}{lccc}
\hline
\textbf{SSL Strategy} & \textbf{Input Feature} & \textbf{Top-1 Acc.} & \textbf{F1-score} \\
\hline

\multirow{3}{*}{Frozen}
 & WavLM        & 88.56 & 87.47 \\
 & wav2vec~2.0  & 87.18 & 85.67 \\
 & HuBERT       & \underline{93.96} & \underline{93.35} \\
\hline

\multirow{3}{*}{Fine-tuned}
 & WavLM        & 89.32 & 88.36 \\
 & wav2vec~2.0  & 87.37 & 85.86 \\
 & HuBERT       & \underline{94.81} & \underline{94.29} \\
\hline

-- & Log-Mel Spec. & \textbf{96.25} & \textbf{95.78} \\
\hline
\end{tabular}
\end{table}

Table~\ref{table:ablation_features} shows that handcrafted log-Mel spectrogram
features achieve the highest performance, reaching a Top-1 accuracy of
$96.25\%$, and outperform all SSL-based representations. Among the SSL features,
HuBERT performs best, improving from $93.96\%$ to $94.81\%$ Top-1 accuracy when
fine-tuning the final transformer layer, while WavLM and wav2vec~2.0
achieve noticeably lower results. This trend is expected, as SSL models are
trained using general speech representation objectives and are not explicitly
optimized for inter-speaker discrimination in closed-set identification.

\subsubsection{Multi-scale temporal encoder}
\label{sec:ablation_temporal_encoder}

\begin{table}[!ht]
\centering
\caption{Ablation study on the multi-scale temporal encoder evaluated on the VoxCeleb1 test set.}
\label{table:ablation_temporal}
\small
\begin{tabular}{lcc}
\hline
\textbf{Temporal Context} & \textbf{Top-1 Acc.} & \textbf{F1-score} \\
\hline
Short-term context  & 90.56 & 89.64 \\
Mid-term context    & 89.96 & 89.10 \\
Long-term context   & 84.90 & 83.51 \\
\hline
Multi-scale temporal context & \textbf{96.25} & \textbf{95.78} \\
\hline
\end{tabular}
\end{table}

Table~\ref{table:ablation_temporal} reports the effect of the temporal context
modeled by the encoder. When using a single temporal stage, the short-term
configuration achieves the highest performance (Top-1 $90.56\%$), followed by
the mid-term configuration (Top-1 $89.96\%$), while the long-term-only setting
yields the lowest performance (Top-1 $84.90\%$). This behavior can be explained
by the fact that many speaker-discriminative characteristics, such as spectral
shape, formant-related information, and phonetic-level articulation patterns,
are primarily captured over short and intermediate temporal contexts and remain
relatively stable across speech segments. In contrast, long-term temporal
modeling alone mainly captures broader utterance-level attributes, such as
speaking rate, rhythm, and global prosodic trends, which are generally less
discriminative when considered in isolation. When long-term context is combined
with short- and mid-term information in TARNet, these complementary cues
reinforce each other, leading to a substantial performance improvement and a
Top-1 accuracy of $96.25\%$.

\begin{table}[!ht]
\centering
\caption{Ablation study on pooling strategies evaluated on the VoxCeleb1 test
set.}
\label{table:ablation_pooling}
\small
\begin{tabular}{lcc}
\hline
\textbf{Pooling Method} & \textbf{Top-1 Acc.} & \textbf{F1-score} \\
\hline
Max Pooling                & 91.13 & 90.25 \\
Temporal Avg. Pooling      & 89.95 & 88.78 \\
Statistics Pooling (SP)    & \underline{95.88} & \underline{95.40} \\
\gls{asp} & \textbf{96.25} & \textbf{95.78} \\
\hline
\end{tabular}
\end{table}

\subsubsection{Attentive Statistics Pooling}
\label{sec:ablation_pooling}

Table~\ref{table:ablation_pooling} shows that simple pooling strategies, such as
max pooling and temporal average pooling, yield noticeably lower performance,
with Top-1 accuracies of $91.13\%$ and $89.95\%$, respectively, as they either
focus on extreme activations or treat all frames equally without modeling frame
importance. Statistics pooling provides a substantial
improvement, increasing Top-1 accuracy to $95.88\%$ by incorporating
second-order statistics that capture utterance-level variability relevant to
speaker identity. The best performance is achieved with attentive statistics
pooling, which further improves Top-1 accuracy to $96.25\%$ by assigning
adaptive frame-level weights, enabling the model to emphasize
speaker-discriminative regions while reducing the influence of less informative or noisy frames.

\begin{figure}[!ht]
    \centering
    \includegraphics[width=\columnwidth]{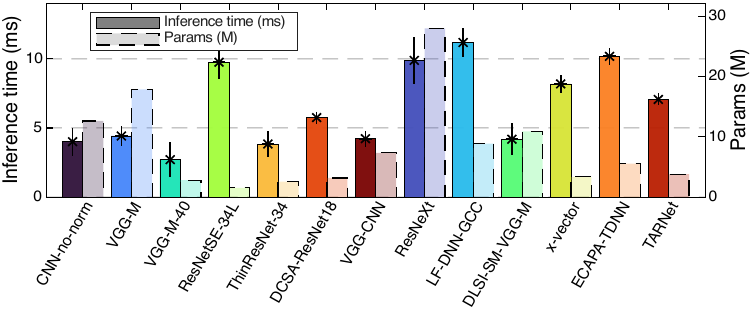}
    \caption{Model complexity comparison in terms of trainable parameters and
    average inference time per utterance.}
    \label{fig:complexity}
\end{figure}

\subsection{Computational Complexity}
\label{sec:complexity}

We compare TARNet with several baseline models in terms of the number of trainable parameters and the average inference time per utterance. The inference time is computed by averaging the per-utterance forward-pass time over all samples in the VoxCeleb1 test set. As shown in Fig.~\ref{fig:complexity}, TARNet contains 3.81M parameters, which is notably fewer than the strong-performing ECAPA-TDNN model (5.56M parameters). In terms of computational efficiency, TARNet achieves an average inference time of 7.07~ms per utterance, which is lower than that of competitive architectures such as ECAPA-TDNN (10.19~ms) and ResNeXt (9.89~ms). These results indicate that TARNet is suitable for real-time and large-scale speaker identification applications.

\section{Conclusion}
\label{sec:conclusion}

This paper presented TARNet, a multi-scale temporal architecture for closed-set speaker identification. TARNet explicitly models short-, mid-, and long-term temporal information and combines these representations through feature fusion and \gls{asp}. Experimental results on VoxCeleb1 and LibriSpeech demonstrate that TARNet consistently outperforms state-of-the-art approaches while maintaining competitive model complexity. Ablation studies further confirm the effectiveness of the proposed components. Future work will investigate the robustness of TARNet under noisy and reverberant conditions and extend the approach to more challenging speaker identification scenarios.

\bibliographystyle{IEEEbib}
\bibliography{refs}

\end{document}

%% file: acronyms.tex
\usepackage[acronym,nonumberlist,style=super,toc]{glossaries}

\newacronym{asr}{ASR}{Automatic Speech Recognition}
\newacronym{cnn}{CNN}{Convolutional Neural Networks}
\newacronym{bilstm}{BiLSTM}{Bidirectional Long Short-Term Memory}
\newacronym{lstm}{LSTM}{Long Short-Term Memory}
\newacronym{mfcc}{MFCC}{Mel Frequency Cepstral Coefficients}
\newacronym{pncc}{PNCC}{Power Normalized Cepstral Coefficients}
\newacronym{gfcc}{GFCC}{Gammatone Frequency Cepstral Coefficients}
\newacronym{sad}{SAD}{Speech Activity Detection}
\newacronym{sir}{SIR}{Signal to Interference Ratio}
\newacronym{snr}{SNR}{Signal-to-Noise Ratio}
\newacronym{rir}{RIR}{Room Impulse Response}
\newacronym{mfb}{MFB}{Mel Filter-Banks}
\newacronym{cat}{CAT-Net}{Channel Attention Temporal Convolutional Network}
\newacronym{fft}{FFT}{Fast Fourier Transform}
\newacronym{tcn}{TCN}{Temporal Convolutional Network}
\newacronym{relu}{ReLU}{Rectified Linear Unit}
\newacronym{prelu}{PReLU}{Parametric ReLU}
\newacronym{osd}{OSD}{Overlapping Speech Detection}
\newacronym{bfi}{BFI}{Backward Frames Inclusion}
\newacronym{crnn}{CRNN}{Convolutional Recurrent Neural Network}
\newacronym{dct}{DCT}{Discrete Cosine Transform}
\newacronym{plp}{PLP}{Perceptual Linear Prediction}
\newacronym{stft}{STFT}{Short-Time Fourier Transform 
}
\newacronym{awgn}{AWGN}{additive White gaussian noise}

\newacronym{gru}{GRU}{Gated Recurrent Unit}
\newacronym{satcn}{SA-TCN}{Self-Attention Temporal Convolutional Network}
\newacronym{vbhmm}{VB-HMM}{Variational Bayes Hidden Markov Model}
\newacronym{vad}{VAD}{Voice activity detection}
\newacronym{hac}{HAC}{Hierarchical Agglomerative Clustering}
\newacronym{der}{DER}{Diarization Error Rate}
\newacronym{jer}{JER}{Jaccard Error Rate}
\newacronym{ar}{AR}{Approximate Randomization}
\newacronym{ssl}{SSL}{ Self-Supervised Learning}
\newacronym{dnn}{DNN}{ Deep Neural Networks}
\newacronym{asp}{ASP}{ Attentive Statistics Pooling}